\documentclass[11pt]{article}
\pdfoutput=1

\usepackage{moriond}

\usepackage{color}
\usepackage{graphicx}

\usepackage{amsmath}

\def\NPB{{\em Nucl. Phys.} B}

\def\PRD{{\em Phys. Rev.} D}

\def\be{\begin{equation}}
\def\ee{\end{equation}}
\def\bea{\begin{eqnarray}}
\def\eea{\end{eqnarray}}

\def\Op{\mathcal{O}}
\newcommand{\lraD}{\stackrel{\mbox{\scriptsize $\leftrightarrow$}}{D}}
\newcommand{\parcite}[1]{$^{[}$\,\cite{#1}$^{\,\!]}$}

\definecolor{darkblue}{cmyk}{1,0.5,0,0.2}

\newcommand{\arXhref}[1]{\href{http://arxiv.org/abs/arXiv:#1}{#1}} 

\newcommand{\Photo}{}

\begin{document}
\hfill SISSA 26/2014/FISI
\vspace*{3.5cm}

\title{PROSPECTS FOR BOUNDS ON ELECTROWEAK AND HIGGS\\ OBSERVABLES VIA SCALING EFFECTS}

\author{ DAVID MARZOCCA}

\address{SISSA, Via Bonomea 265, 34143 Trieste, Italy}

\maketitle\abstracts{
In these proceedings we briefly review the basic concepts underlying indirect bounds on the Wilson coefficients of some Standard Model dimension-6 operators, relevant to electroweak and Higgs observables, obtained via renormalization group (RG) mixing to strongly constrained observables. With the present data we derive RG-induced bounds, stronger than the direct constraints, on some Higgs couplings and anomalous triple gauge couplings. Any deviation from these bounds would suggest a particular pattern of correlations among the Wilson coefficients, thus offering a new window on the new physics sector. Prospects for these effects at the LHC and at proposed future lepton colliders (ILC and TLEP) are assessed.}

\section{Introduction}

A new physics sector at the TeV scale, related to the electroweak (EW) sector of the Standard Model (SM), is well motivated by the instability of the EW scale under quantum corrections.
The absence of evidence for such new physics (NP) motivates the assumption that all new particles have masses much bigger than the EW scale, in which case an efficient and model-independent way to parametrize deformations from the SM is in the context of effective field theories, that is adding to the SM action effective operators with scaling dimension bigger than four and invariant under the SM gauge symmetries \parcite{Buchmuller:1985jz}.
Assuming baryon and lepton number conservation at the NP scale $\Lambda \sim$ TeV, the leading deformations are described by dimension 6 operators invariant under the SM gauge group, $\delta \mathcal{L} = \sum_i c_i / \Lambda^2 ~ \Op_i$. 
It has been shown that, for one family of fermions, there are 59 independent operators \parcite{Grzadkowski:2010es}.

These operators are generated at the scale $\Lambda$ upon integrating out the heavy degrees of freedom; their effect is instead measured at the electroweak scale $\sim m_W$.
Among these two energy scales the Wilson coefficients follow the renormalization group (RG) flow and, in particular, mix among themselves.
This mixing opens the possibility to link different deformations which would be, otherwise, unrelated \parcite{Hagiwara:1993ck}. Our aim is to employ this link in order to derive RG-induced bounds, from the one-loop mixing, which are stronger than the direct tree-level constraints, under an assumption on the UV dynamics that we want to test.

\section{EW and Higgs observables}

For a completely general analysis all operators should be considered, however a set of observables will receive contributions only from a subset of operators.
Given that the new sector should cure the quantum instability of the Higgs mass, we expect that some of the most important deformations from the SM will involve the Higgs sector. For this reason we focus on a set of ten EW and Higgs pseudo-observables, particularly relevant for universal new physics scenarios:
the EW oblique parameters $\hat{S},\hat{T},W$ and $Y$; the three independent anomalous triple gauge couplings (TGC) $g_1^Z, k_\gamma, \lambda_\gamma$; the Higgs couplings to two photons ($\hat{c}_{\gamma\gamma}$) and to $Z \gamma$ ($\hat{c}_{\gamma Z}$) and a universal rescaling of the Higgs branching ratios ($\hat{c}_{H}$). For the precise definition of these observables we refer to ref.~\parcite{Elias-Miro:2013eta}.

{\renewcommand{\arraystretch}{1.4} 
\begin{table}[tc]
\small
\centering
\begin{tabular}{ccc}
\begin{tabular}{|c|}\hline
${\cal O}_H=\frac{1}{2}(\partial^\mu |H|^2)^2$\\
${\cal O}_T=\frac{1}{2}\left (H^\dagger \lraD_\mu H\right)^2$\\
${\cal O}_W=ig\left( H^\dagger  \tau^a \lraD^\mu H \right )D^\nu  W_{\mu \nu}^a$\\
${\cal O}_B=ig'  Y_H \left( H^\dagger \lraD^\mu H \right )\partial^\nu  B_{\mu \nu}$\\
${\cal O}_{2W}=-\frac{1}{2}  ( D^\mu  W_{\mu \nu}^a)^2$\\
${\cal O}_{2B}=-\frac{1}{2}( \partial^\mu  B_{\mu \nu})^2$\\ \hline 
\end{tabular}
&\qquad \qquad&
\begin{tabular}{|c|}\hline
${\cal O}_{BB}=g^{\prime 2} |H|^2 B_{\mu\nu}B^{\mu\nu}$\\
${\cal O}_{WB}= g{g}^{\prime} H^\dagger \sigma^a H W^a_{\mu\nu}B^{\mu\nu}$\\
${\cal O}_{WW}=g^2 |H|^2 W^a_{\mu\nu}W^{a \mu\nu}$\\
${\cal O}_{3W}= \frac{1}{3!} g\epsilon_{abc}W^{a\, \nu}_{\mu}W^{b}_{\nu\rho}W^{c\, \rho\mu}$\\ \hline
 \end{tabular}
\end{tabular}
\caption{\small The 10 CP-even operators made of SM bosons that provide the leading contributions to the selected set of observables.
\label{table:operators}}
\vspace{-0.2cm}
\end{table}}
The subset of ten operators which most efficiently parametrize these observables is listed in table~\ref{table:operators}; the complete operator basis of which these operators are part of, is defined in ref.~\parcite{Elias-Miro:2013eta}.
In order to take make more direct the connection with the selected set of observables,
we go to a new basis in which to each observable corresponds only one coefficient, $\delta (\text{obs})_i = \hat{c}_i$, that we call \emph{observable coefficient}:
\be \begin{array}{c}
	\hat{T}=\hat{c}_T (m_W)~, \quad \hat{S}=\hat{c}_S (m_W)~, \quad Y=\hat{c}_{Y}(m_W)~, \quad W=\hat{c}_{W}(m_W), \\
	\delta g_1^Z = \hat{c}_{gZ}  (m_W)~, \quad \delta \kappa_\gamma = \hat{c}_{\kappa\gamma} (m_W)~, \quad \lambda_Z = \hat{c}_{\lambda \gamma} (m_W)~, \\
	\hat{c}_{\gamma\gamma}~, \quad \hat{c}_{\gamma Z}~, \quad \hat{c}_{H}.
	\label{eq:ObsCoeff}
\end{array}\ee
In this coefficients we also include the $\Lambda$ dependence: $\hat{c} \sim c ~ m_W^2 / \Lambda^2$. The simple relations with the Wilson coefficients of the operators in table~\ref{table:operators} can be found in ref.~\parcite{Elias-Miro:2013eta}.

\section{RG-induced bounds}

At leading-log order, the measurable low-energy Wilson coefficients are related to the UV ones via
\be
	\delta (\text{obs})_i|_{m_h} = \hat{c}_i(m_h) =  \hat{c}_i(\Lambda)-\frac{1}{16\pi^2} \hat{\gamma}_{ij}\hat{c}_j(\Lambda) \log\left(\frac{\Lambda}{m_h}\right) \ ,
\label{ObsCoeffRunning}
\ee
where $\hat{\gamma}_{ij}$ is the anomalous dimension matrix. In ref.~\parcite{Elias-Miro:2013eta} we computed the relevant sub matrix for the operators in table~\ref{table:operators} (the full matrix can be found in ref.~\parcite{Alonso:2013hga}, computed in the operator basis of ref.~\parcite{Grzadkowski:2010es}, which makes the relation to the set of observables in eq.~\eqref{eq:ObsCoeff} not straightforward).
Assuming absence of tuning, or correlations, in eq.\eqref{ObsCoeffRunning} among the UV coefficients, we require each addend on the r.h.s. to be bounded by the same constraint as the l.h.s. coefficient. Given the wide difference in precision of the experimental constraints on these observables and some accidentally big numerical factors in the anomalous dimensions, some of the RG-induced bounds obtained in this way are stronger than the direct ones. This allows to start probing the tuning (i.e. the correlations) among the UV coefficients, thus offering a new window on the UV dynamics.

\begin{figure}[t]
\begin{center}
\hspace*{-0.65cm} 
\begin{minipage}{0.5\linewidth}
\begin{center}
	\hspace*{0cm} 
	\includegraphics[width=\linewidth]{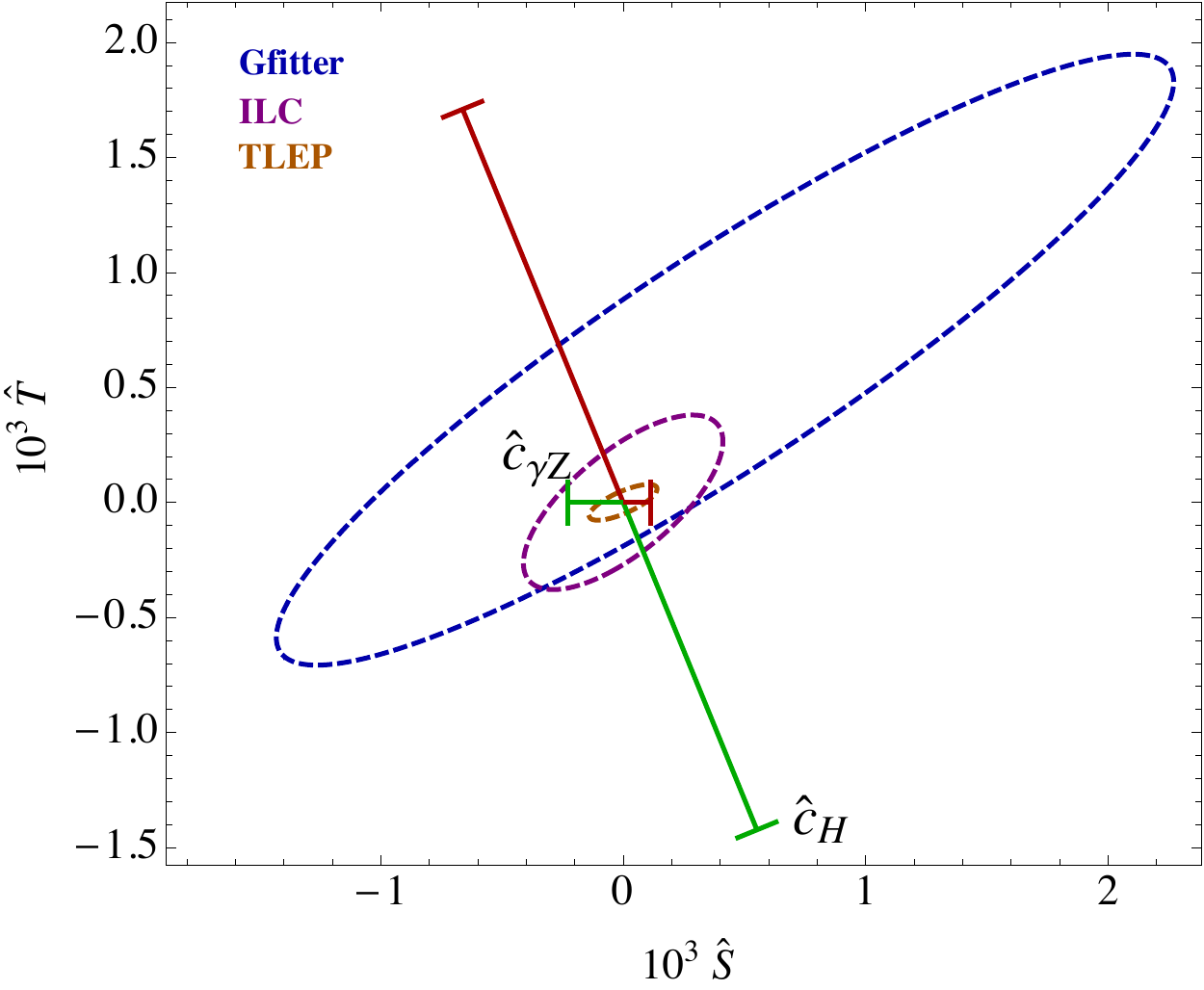}
\end{center}
\end{minipage}
%
%
\begin{minipage}{0.5\linewidth}
\begin{center}
	\hspace*{0cm} 
	\includegraphics[width=\linewidth]{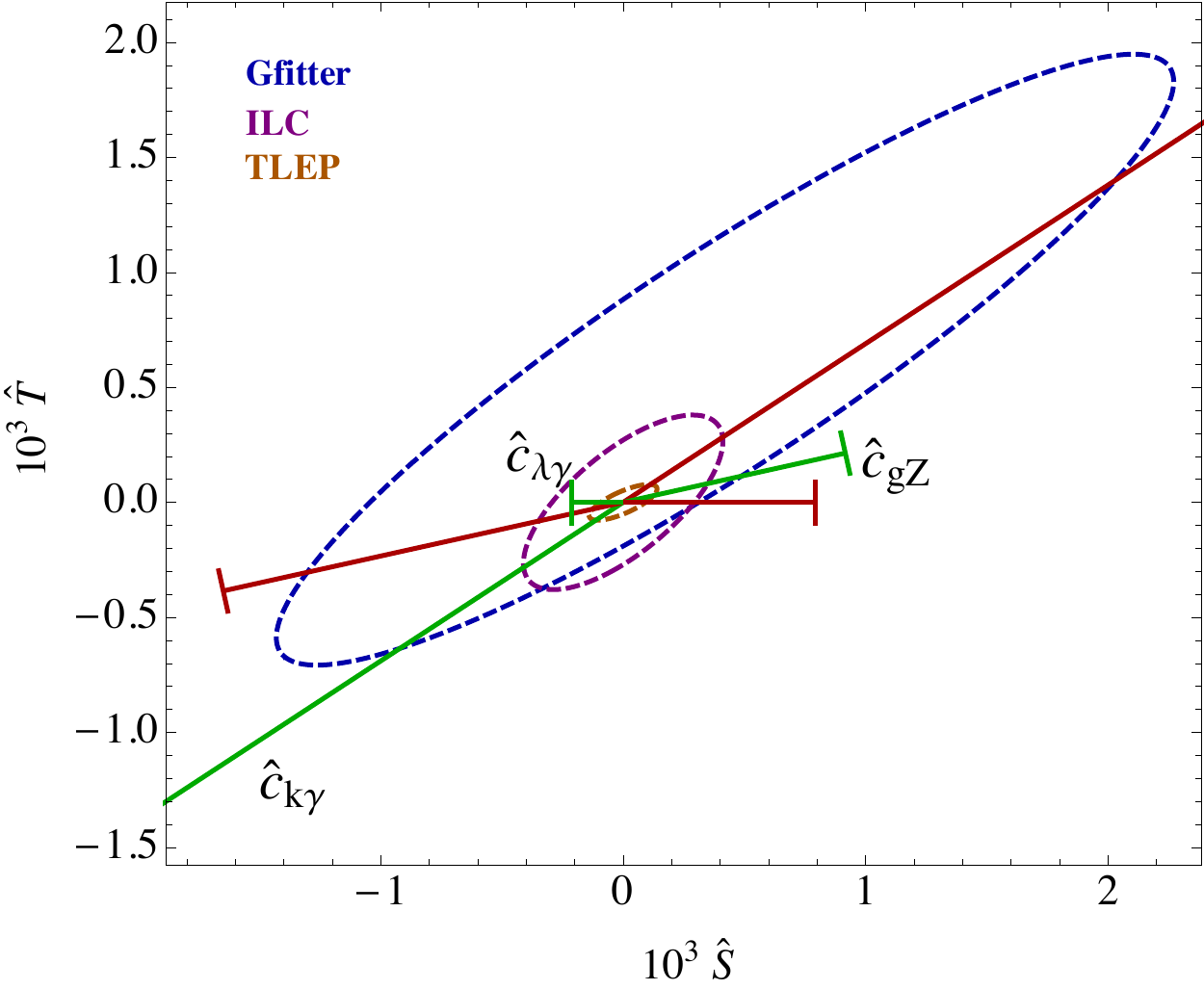}
\end{center}
\end{minipage}
\end{center}
\caption{\small The dashed ellipses represent the 95\% CL bound on $\hat{S}$ and $\hat{T}$ at present (blue) and predicted for the ILC (purple) and TLEP (orange).
The straight lines represent the RG-induced contribution to the oblique parameters from the weakly constrained observable couplings, divided in Higgs couplings (left panel) and TGC couplings (right panel), for $\Lambda = 2 \mbox{ TeV}$. The length of the lines corresponds to their present 95\% CL direct bounds; the line is green (red) for positive (negative) values of the parameters.
\label{fig:RGBounds}}
\vspace{-0.2cm}
\end{figure}

At present, the strongest RG-induced bounds are obtained by considering the RG contributions to the $\hat{S}$ and $\hat{T}$ parameters. Using the $\hat{S}-\hat{T}$ ellipse of ref.~\parcite{gfitter}, we present the RG-induced bounds in figure~\ref{fig:RGBounds}, for a reference value $\Lambda = 2$ TeV. For each of the observable couplings in eq.\eqref{eq:ObsCoeff} constrained directly at the percent level, or worse, we obtain a RG-induced bound (table~\ref{tab:RGST}) which is stronger than, or of the same order as, the direct tree-level bound (table~\ref{tab:future}).
Another strongly constrained observable coefficient is $\hat{c}_{\gamma \gamma}$. The only coefficients mixing to this observable at one-loop, among the ones we considered, are the two anomalous TGC $\hat{c}_{k\gamma}$ and $\hat{c}_{\lambda \gamma}$\parcite{Elias-Miro:2013eta}. The RG-induced bounds obtained in this way are shown in the second column of table~\ref{tab:RGgammagamma}.
Even though at present these are not competitive with the ones coming from the mixing to the $\hat{S}$ and $\hat{T}$ parameters, in the future they will provide a strong handle for searching correlations in the UV dynamics, given the very good precision expected for the measurement of $\hat{c}_{\gamma \gamma}$.

\section{Future prospects and tuning}

Let us now discuss the future prospects for the RG-induced bounds, given the expected sensitivities on the observable couplings introduced above for $300$ fb$^{-1}$ and $3000$ fb$^{-1}$ \parcite{aTGCLHC} of luminosity at the LHC and for the ILC \parcite{ILC} and TLEP \parcite{TLEP} projects, as collected in table~\ref{tab:future}.

\begin{table}[b]
\small
\begin{center}
\begin{tabular}{l | c|c|c|c|c}
Obs.	& Now & LHC (300 fb$^{-1}$)  & HL-LHC (3 ab$^{-1}$) & ILC & TLEP \\ \hline
$\hat{c}_S$  &  $[-1,2] \times 10^{-3}$~\parcite{gfitter}  &  --  &  --  &  $1.4 \times 10^{-4}$~\parcite{Baak:2013fwa}  &  $5 \times 10^{-5}$~\parcite{Mishima:2013} \\
$\hat{c}_T$  &  $[-1,2] \times 10^{-3}$~\parcite{gfitter}  &  --  &  --  &  $1.6 \times 10^{-4}$~\parcite{Baak:2013fwa}  &  $3.1 \times 10^{-5}$~\parcite{Mishima:2013} \\
$\hat{c}_{gZ}$  &  $[-4,2] \times 10^{-2}$~\parcite{leptgc}  &  $3 \times 10^{-3}$~\parcite{aTGCLHC}  &  $2 \times 10^{-3}$~\parcite{aTGCLHC}  &  $1.8 \times 10^{-4}$~\parcite{ILC}  &  n.a. \\
$\hat{c}_{k\gamma}$  &  $[-10,7] \times 10^{-2}$~\parcite{leptgc}  &  $3 \times 10^{-2}$~\parcite{aTGCLHC}  &  $1 \times 10^{-2}$~\parcite{aTGCLHC}  &  $1.9 \times 10^{-4}$~\parcite{ILC}  &  n.a. \\
$\hat{c}_{\lambda \gamma}$  &  $[-6,2] \times 10^{-2}$~\parcite{leptgc}  &  $9 \times 10^{-4}$~\parcite{aTGCLHC}  &  $4 \times 10^{-4}$~\parcite{aTGCLHC}  &  $2.6 \times 10^{-4}$~\parcite{ILC}  &  n.a.  \\
$\hat{c}_{\gamma \gamma}$  &  $[-1,2] \times 10^{-3}$~\parcite{Pomarol:2013zra}  &  $1 \times 10^{-4}$~\parcite{Dawson:2013bba}  &  $4 \times 10^{-5}$~\parcite{Dawson:2013bba}  &  $7.6 \times 10^{-5}$~\parcite{Dawson:2013bba}  &  $2.9 \times 10^{-5}$~\parcite{Dawson:2013bba,TLEP}  \\
$\hat{c}_{\gamma Z}$  &  $[-6,10] \times 10^{-3}$~\parcite{Pomarol:2013zra}  &  $9 \times 10^{-4}$~\parcite{Dawson:2013bba}  &  $2 \times 10^{-4}$~\parcite{Dawson:2013bba}  &  n.a.  &  n.a. \\
$\hat{c}_{H}$  &  $[-6,5] \times 10^{-1}$~\parcite{Pomarol:2013zra}  &  $1 \times 10^{-1}$~\parcite{Dawson:2013bba}  &  $5 \times 10^{-2}$~\parcite{Dawson:2013bba}  &  $5 \times 10^{-2}$~\parcite{Dawson:2013bba}  &  $1 \times 10^{-2}$~\parcite{Dawson:2013bba,TLEP}
\end{tabular}
\end{center}
\caption{\small Future prospects in the direct determination of the observable couplings discussed here from the LHC, a high-luminosity LHC, the ILC at 800GeV and from TLEP after a first phase at 240GeV and a second one at 350GeV. The precision in $\hat{S}, \hat{T}$ will not improve sensibly at the LHC or HL-LHC and the other missing elements have not yet been studied in the literature.}
\label{tab:future}
\end{table}

The precision on the oblique parameters could reach the $10^{-4}$ level at ILC \parcite{Baak:2013fwa} and the $10^{-5}$ level at a TLEP collider \parcite{Mishima:2013}. This would allow to improve sensibly the RG-induced bounds on our set of observable couplings, as can be seen in figure~\ref{fig:RGBounds} and in table~\ref{tab:RGST}

The measurement of the Higgs couplings, in particular the one to two photons $\hat{c}_{\gamma\gamma}$, will improve substantially in the future: by one order of magnitude at 14TeV LHC with 300 fb$^{-1}$ of integrated luminosity and at the ILC, and almost two orders of magnitude at a high-luminosity LHC phase and at a TLEP collider \parcite{Dawson:2013bba}. The prospects for RG-induced bounds on the observable coefficients which mix to $\hat{c}_{\gamma \gamma}$ are reported in table~\ref{tab:RGgammagamma}.

\begin{table}[t]
\small
\begin{center}
\begin{tabular}{l | c|c|c}
mix.	to $(\hat{S},\hat{T})$	& Now &  ILC & TLEP \\ \hline
$\hat{c}_{\gamma Z}$  &  $[-2,6] \times 10^{-2}$  &  $2 \times 10^{-2}$  &  $5 \times 10^{-3}$ \\
$\hat{c}_{H}$  &  $[-2,0.5] \times 10^{-1}$  &  $7 \times 10^{-2}$  &  $2 \times 10^{-2}$ \\
$\hat{c}_{gZ}$  &  $[-3,1] \times 10^{-2}$  &  $8 \times 10^{-3}$  &  $3 \times 10^{-3}$ \\
$\hat{c}_{k\gamma}$  &  $[-5, 2] \times 10^{-2}$  &  $9 \times 10^{-3}$  &  $3 \times 10^{-3}$ \\
$\hat{c}_{\lambda \gamma}$  &  $[-2,8] \times 10^{-2}$  &  $2 \times 10^{-2}$  &  $7 \times 10^{-3}$
\end{tabular}
\end{center}
\caption{\small Present status and future prospects for the RG-induced bounds, for $\Lambda = 2$ TeV, from the mixing to $(\hat{S},\hat{T})$, given the predicted sensitivity in this observables at ILC and TLEP, as shown in table~\ref{tab:future}.}
\label{tab:RGST}
\end{table}

\begin{table}[t]
\small
\begin{center}
\begin{tabular}{c | c|c|c|c|c}
mix.	to $\hat{c}_{\gamma \gamma}$ & Now  &  LHC & HL-LHC & ILC & TLEP \\ \hline
$\hat{c}_{k\gamma}$  &  $[-0.2,0.3]$  &  $2 \times 10^{-2}$  &  $7 \times 10^{-3}$  &  $1 \times 10^{-2}$ &  $5 \times 10^{-3}$ \\
$\hat{c}_{\lambda \gamma}$  &  $[-0.05,0.10]$  &  $5 \times 10^{-3}$  &  $2 \times 10^{-3}$  &  $4 \times 10^{-3}$  &  $1 \times 10^{-3}$ \\ 
\end{tabular}
\end{center}
\caption{\small Present status and future prospects for the RG-induced bounds, for $\Lambda = 2$ TeV, on two anomalous TGC from the mixing to $\hat{c}_{\gamma \gamma}$, given the predicted sensitivity in this observable as shown in table~\ref{tab:future}.}
\label{tab:RGgammagamma}
\vspace{-0.2cm}
\end{table}

If a deviation from the SM will be observed (i.e. one observable coefficient will have a direct bound $0 < \epsilon_j^{low} < |\hat{c}_j(m_W)| < \epsilon_j^{up}$), by comparing the lower bound $\epsilon_j^{low}$ with the RG-induced bound on $\hat{c}_j$ ($|\hat{c}_j| < \epsilon_{ji}^{RG}$) obtained considering its RG mixing to a strongly constrained observable $\hat{c}_i$ (like $\hat{S}$ and $\hat{T}$), we can determine the necessary amount of tuning in eq.\eqref{ObsCoeffRunning}.
By taking the logarithmic derivative of eq.\eqref{ObsCoeffRunning} with respect to the UV coefficient $\hat{c}_j(\Lambda)$ one gets that the tuning is \parcite{Elias-Miro:2013eta} $\Delta_{ij} > \epsilon_j^{low} / \epsilon_{ji}^{RG} $. 
Therefore, if $\epsilon_j^{low} \gg \epsilon_{ji}^{RG}$ a definite amount of tuning (or of correlation) in the UV dynamics would be necessary. This could provide a new window on the UV physics.

For example, if $\hat{c}_H$ should be measured to be $\sim 0.2$ (0.1) while no deviation in $(\hat{S},\hat{T})$ should be observed after TLEP, the RG-induced bound $|\hat{c}_H^{RG,\text{TLEP}}| < 2 \times 10^{-2}$ would imply a tuning $\Delta_{H, (S,T)} > 10$ (5).
Similarly, should one measure $\hat{c}_{k \gamma} \sim 5 \times 10^{-2}$, the RG-induced bound from $(\hat{S},\hat{T})$ at TLEP, $|\hat{c}_{k \gamma}^{RG,\text{TLEP}}| < 3 \times 10^{-3}$, would imply a tuning $\Delta_{k_\gamma, (S,T)} > 17$.

\vspace{-0.1cm}
\section*{Acknowledgements}
\vspace{-0.2cm}
I am grateful to J. Elias-Mir\'o, C. Grojean and R. S. Gupta for the stimulating collaboration and for providing precious comments on this manuscript. I also thank the Moriond organizers for the grant award to attend the conference and the opportunity to present this work.

\end{document}